\documentclass[aps,prl,showpacs,twocolumn,groupedaddress,amssymb]{revtex4}

\usepackage{graphicx}
\usepackage{dcolumn}
\usepackage{bm}

\begin{document}

\title{Cyclotron Resonance in Strongly Magnetized Plasmas and Gamma Ray Burst}

\author{S. Son}
\affiliation{18 Caleb Lane, Princeton, NJ 08540}
\author{Sung Joon Moon}
\affiliation{28 Benjamin Rush Lane, Princeton, NJ 08540}
\date{\today}

\begin{abstract}
A plausible scenario for the gamma ray and the hard x-ray burst in a strongly magnetized plasma,
based on the collective plasma maser instability, is proposed.
The physical parameters with which this scenario becomes relevant are estimated.
The attractive feature of this scenario over the conventional cyclotron radiation theory is discussed. 
\end{abstract}

\maketitle
\section{Introduction}
According to a recent study, 
the thermal electron 
gyro-motion in a strongly magnetized plasma could lead to instabilities
of  electromagnetic (E\&M) waves~\citep{sonmaxwell}.
This instability, so-called the maser instability, has significant implications 
on strongly magnetized plasmas~\citep{Stone,Biskamp, Innes,Fishman, Nakar, Horse}.
The goal of this paper is to estimate its relevance to the gamma ray or hard x-ray burst
in the strongly magnetized astrophysical plasmas~\citep{Fishman, Nakar}.
The gamma ray burst in the absence of the magnetic field~\citep{Mendonça} 
is not relevant to our analysis. 
Our analysis shows that the photons of 10 keV to 1 MeV can be emitted from a relativistic plasma 
with the electron density $ 10^{19} - 10^{26}~\mathrm{cm^{-3}} $ and the magnetic field
of order of $10^{8}$ to $10^{13}$ gauss.  
In contrast to the conventional cyclotron radiation 
where the photons are emitted rather uniformly,
the angular distribution of the radiated photons could be sharply concentrated. 
Advantages of this possible phenomenon over the ones based on the incoherent 
cyclotron radiation are discussed.

The rest of the paper is organized as follows. 
The instability growth rate is presented in Sec. 2, based on the Landau damping theory,
which is analyzed when the photon wave vector is parallel to the magnetic field (Sec. 3)
and when the photon wave vector is not parallel to the magnetic field (Sec. 4).
The gamma ray burst is discussed in Sec. 5, and the difference between this theory and
the conventional theories is presented Sec. 6, which is followed by summary (Sec. 7).

\section{Coherent Cyclotron Radiation}
Let us consider the growth rate of a collective E\&M wave
in the presence of a strong magnetic field $B_0\hat{z}$;
the detailed derivation when an E\&M wave propagates parallel to the magnetic field
is in Ref.~\citep{sonmaxwell}.
The physical mechanism of the instability is similar to that of the maser or
the free electron laser, where a coherent E\&M wave gets amplified by the electron
gyro-motion as manifested in the radio frequency wave of the solar corona~\citep{chu,Treumann}. 
The instability regime is ubiquitous in strongly magnetized plasmas (\citep{sonmaxwell}).
To summarize briefly, the kinetic motion of a relativistic electron is solved
in the presence of an E\&M mode to the second order, and the ensemble average of
the electron kinetic loss (gain) is taken over the electron distribution.
Equating this energy loss to the E\&M wave growth,
the E\&M growth (decay) rate is obtained.
The growth rate for a general propagation direction $\theta$,
which is defined to be the angle between the magnetic field and the photon wave-vector, is
\begin{eqnarray} 
\Gamma  = &+& \frac{1}{\zeta} \left[\frac{\pi}{2} \frac{\omega_{pe}^2}{c^2k^2} 
\langle \frac{\Omega_1(\omega, k, \theta, \mathbf{\beta})}{\gamma}\frac{\mathbf{\nabla_{\beta}} S \cdot\mathbf{ \nabla_{\beta}} f}{|\nabla S(\mathbf{\beta})|^2 } \frac{\beta_{\perp}^2}{2}     \rangle_{S=0} \right]  \nonumber \\
 &-& \left[\frac{\pi}{2} \frac{\omega_{pe}^2}{c^2k^2}   
\langle  \frac{\Omega_2(\omega, k, \theta, \mathbf{\beta}) }{\gamma} f \rangle_{S=0}\right] \mathrm{,} \label{eq:landau2} \\  \nonumber 
\end{eqnarray}
where  $k$ ($\omega$) is the wave vector (frequency) of the E\&M mode, 
$\mathbf{\beta} = (\beta_{\perp}, \beta_{\parallel}) = \mathbf{v} / c$,
$\beta_{\perp}^2 = \beta_x^2 + \beta_y^2 $, 
$\omega_{pe}^2 = 4 \pi n_e e^2 /m_e$ is the plasma frequency,  
$S(\mathbf{\beta}) = \beta_{\parallel}\cos(\theta) - \omega / ck  + \omega_{ce} / (ck \gamma(\beta))$, $\omega_{ce} = eB_0/m_e c$ is the classical cyclotron frequency,  
$\gamma(\beta) = (1 - \beta^2)^{-1/2}$, $f$ is the electron distribution with the normalization
of   $\int f d^3 \mathbf{\beta} = 1$, 
$\langle A \rangle_{S=0} = \int \delta(S) A d^3 \mathbf{\beta} $ 
is the integration in the velocity space with the constraint $S=0$,
and $\Omega_1$ and $\Omega_2$ are obtained below.
Defining $\zeta$ to be the ratio of the wave energy density 
to the wave energy density in vacuum, $ \zeta = E_w/( E_x^2/4\pi)$, 
it is assumed in our analysis that $\zeta > 0$, which is the case for
most of E\&M waves in dense plasmas. 
For a generic angle $\theta$, there could be two independent modes; TE and TM modes.
The wave vector is given as $\mathbf{k} = k\cos\theta \hat{z} + \sin\theta \hat{x}$.   
Let us define the TE (TM) mode as 
$\mathbf{E} = E_1\cos(\mathbf{k} \cdot \mathbf{r} - \omega t) \hat{x} $ 
and 
$\mathbf{B} = E_1 (ck/\omega)\cos(\mathbf{k} \cdot \mathbf{r} - \omega t)  (\cos\theta \hat{y} -  \sin\theta \hat{z} )$ 
($\mathbf{E} = E_1 \cos(\mathbf{k} \cdot \mathbf{r} - \omega t) (\cos\theta \hat{x} -\sin\theta \hat{z} )$ and $\mathbf{B} = E_1 (ck/\omega) \cos(\mathbf{k} \cdot \mathbf{r} - \omega t) \hat{y}$).    
$\Omega_1$ and $\Omega_2$ for the TE mode are
 \begin{eqnarray} 
\Omega_1(\theta) &=& \frac{c^2k^2}{\omega}\cos\theta \left(\cos\theta- \frac{ck}{\omega}\beta_{\parallel}\right)  - \frac{\omega_{ce}}{\gamma} \nonumber  \\
\Omega_2(\theta) &=& ck \left(1-\frac{\beta_{\perp}^2}{2} -\frac{ck}{\omega} \beta_{\parallel}\cos\theta\right) \mathrm{.} \label{eq:te}
 \end{eqnarray}
For the TM mode, they are given as 
 \begin{eqnarray} 
\Omega_1(\theta) &=& \frac{c^2k^2}{\omega}\cos\theta \left(1 - \frac{ck}{\omega}\beta_{\parallel}\cos\theta\right)  - \frac{\omega_{ce}}{\gamma}\cos\theta \nonumber  \\
\Omega_2(\theta) &=& ck \left((1-\frac{\beta_{\perp}^2}{2})\cos\theta -\frac{ck}{\omega} \beta_{\parallel}\right) \mathrm{.} \label{eq:tm}
\end{eqnarray}
The non-relativistic limit of Eq.~(\ref{eq:landau2}) is
\begin{equation} 
\Gamma = \frac{\pi}{2} \frac{1}{\zeta} \frac{\omega_{pe}^2}{k^2} 
\left( \langle \frac{ \beta_{\perp}^2}{2}  \Omega_1(\theta) \frac{d f}{d v}\rangle 
-  \langle \frac{f(v)}{c} \Omega_2(\theta)\rangle \right)  \label{eq:landau3}
   \mathrm{,}
\end{equation}
where
$\langle \rangle$ is the ensemble average with $v_z = v_r =(\omega-\omega_{ce})/k$ or $\langle A \rangle =  \int   A\delta(v -v_r) d^3 v $. 
When $\theta=0$, the Eqs.~(\ref{eq:landau2}) and (\ref{eq:landau3}) are reduced to
the previously obtained results~\citep{sonmaxwell}.

For more rigorous treatment when $\theta \neq 0 $, there are infinite expansions of
the Bessel function $J_n(k \cos (\theta) r_g)$, where $r_g$ is the electron gyro-radius.
However, the unstable wave identified in our analysis has the condition $k\cos(\theta) r_g < 1$
so that $J_1(k \cos (\theta) r_g) \cong 1$. 
For this reason, Eq.~(\ref{eq:landau2}) is derived assuming $J_1 = 1$. 

\section{Case where $\theta=0$ and $\omega \neq ck$}

Consider the case where the E\&M field propagates parallel with the magnetic field ($\theta=0$).
There is no instability  if  $\omega  = ck$, because $\Omega_1 = 0$. 
For simplicity, we refer to the first (second) term on the right-hand side of
Eqs.~(\ref{eq:landau2}) and (\ref{eq:landau3}) as the gyro-lasing (gyro-damping) term.  
For $\omega > ck$, the semi-classical instability growth rate is shown to be~\citep{sonmaxwell}
\begin{equation} 
\Gamma =\frac{\pi}{2}  \frac{1}{\zeta}\frac{\omega_{pe}^2}{c^2k^2}
\left( \frac{v_r}{ c^2 } \Omega_1  - (1-\beta_{the}^2 - \frac{ck}{\omega}\beta_r) k\right) f_{\parallel}(v_r) \mathrm{,}   \label{eq:cla}
\end{equation}
where $\beta_{the} = v_{the} / c $, $\beta_r= v_r/c = (\omega - \omega_{ce})/c$,
$v_{the}$ is the electron thermal velocity, and $f_{\parallel}(v_r) = \int f \delta(v_z - v_r) d^3 \mathbf{v}$. 
The instability criterion is $\Gamma > 0 $.  
The maximum possible instability, ignoring the gyro-damping term,  
is  estimated to be
\begin{equation} 
\Gamma_{max} \cong  0.19 \times  \frac{\omega_{\mathrm{pe}}^2}{ (ck)^2} \Omega_1 \mathrm{,} \label{eq:max}
\end{equation}
where it is assumed that $\zeta\cong 1$~\citep{sonmaxwell}.

The Maxwellian distribution for fully relativistic electrons is
$f(\beta) \cong \gamma^3 \exp(- \gamma \lambda) $, where 
$\lambda =  m_e c^2 / T_e$.  
This is so-called the Maxwell-Juttner's distribution~\citep{jutt}. 
When $\lambda < 1$, 
the distribution  is peaked at $\gamma  = 3/\lambda$ 
with the width of $\delta \beta \cong (\lambda/3)^2$. 
We consider the case
when $\omega > ck >\omega_{ce}$.
Assuming $\beta \cong 1 $, the condition $\Gamma > 0 $ can be re-casted as 
\begin{eqnarray}
\int_{S=0} d^3\mathbf{\beta } \left(\frac{\mathbf{\nabla_{\beta}} S \cdot\mathbf{ \nabla_{\beta}} f}{|\nabla S(\mathbf{\beta})|^2 }\right)  \left(\frac{\omega_{ce} }{ \gamma ck } + 1\right) \left(1-\frac{ck}{\omega} \right) \nonumber \\ \nonumber   \\ 
>   \int_{S=0} d^3\mathbf{\beta }   f(\beta_{res})\left(1-\frac{ck}{\omega}\beta_{res} \right) \mathrm{.} \label{eq:inst2} 
\end{eqnarray}
One necessary condition for the existence of the resonance is given as 
 $ \omega / ck < \beta_{max} + \omega_{ce}/\gamma_{max} = 
\sqrt{1 + \omega_{ce}^2/ c^2k^2}$, 
where  $\gamma_{max} = (1 -\beta_{max}^2)^{-1/2}$ and  $\beta_{max} = 1/ \sqrt{ 1 + (\omega_{ce}/ck)^2 }$. 
If the electron distribution has the relatively high slope near  the resonance region ($\gamma_{max} \cong  3 T_e / m_e c^2$), 
Eq.~(\ref{eq:inst2}) is  possible since  $ |\nabla_{\beta} S | \ll 1$ near the resonance. 
The growth rate without the gyro-damping term can be estimated, using 
$\nabla_{\beta} f \cong \gamma^2(3-\lambda \gamma) f \mathbf{\beta}$   and $\beta \cong 1$, as  
\begin{equation}  \Gamma_{max} \cong  \frac{\omega_{\mathrm{pe}}^2}{ (ck)^2}
\int \delta(S) \left[\gamma \Omega_1 (3-\lambda) f \frac{dS/d\beta   }{|\nabla_{\beta}S|^2}\right] d^3 \mathbf{\beta}   \mathrm{, } \label{eq:max2}
\end{equation} 
where  $dS/d\beta = (\mathbf{\beta} \cdot \nabla S)/|\mathbf{\beta}|$.   
\section{Case where $\theta \neq 0$ and $\omega = ck$}  
 
The resonance condition for the TM mode is
$S= \beta_{\parallel} \cos\theta + \omega_{ce} / ck \gamma - 1 = 0$. 
The gyro-lasing term vanishes at the resonance 
because $\Omega_1 =( c^2k^2/\omega) S \cos(\theta) =0$  from Eq.~(\ref{eq:tm}). However,
the gyro-damping term does not vanish ($\Omega_2 / ck =  \cos\theta(1-\beta_{\perp}^2/2) - \beta_{\parallel}$). 
If $\Omega_2 < 0$ at $S=0$, 
this term acts as an amplifying term instead of a damping term. 
Consider the semi-classical case. 
From the resonance condition $\beta_{r} \cos\theta = 1 - \omega_{ce}/ck$, 
$\Omega_2$ is given by
\begin{equation} 
   \Omega_2= - ck \left[ \beta_r - \frac{(1-\frac{\omega_{ce}}{ck})(1-\frac{\beta_{\perp}^2}{2})}{\beta_r }\right] \mathrm{.}
\end{equation}
The instability condition ($\Omega_2 > 0$)  is 
given as  $\beta_r^2 > (1- \omega_{ce}/ck)(1-\beta{\perp}^2/2)$. 
If $\beta_{\perp} \cong 0$,  the condition is reduced to 
 $ \beta_r >  \sqrt{1- \omega_{ce}/ck} $, or
 $\cos\theta = (1-\omega_{ce}/ck)/\beta_r <  \sqrt{1- \omega_{ce}/ck}$. 
For a given Maxwellian distribution $f_M$ with the normalization $\int f d \beta_z = 1$,
the growth rate is
\begin{equation} 
\Gamma = \frac{\pi}{2}  \frac{1}{\zeta} \frac{\omega_{pe}^2}{c^2k^2} ck 
 \left[ \beta_r -   \frac{(1-\frac{\omega_{ce}}{ck})(1-\frac{\beta_{\perp}^2}{2})}{\beta_r }\right] f_M(\beta_r) \mathrm{.}
\end{equation}
The growth rate has the maximum  similar to  Eq.~(\ref{eq:max})
when $\beta_{the} < \beta_r < 2 \beta_{the}$.
For given $\omega_{ce}$ and $ck$,   the growth rate as a function of $\theta$
has the maximum when $ \beta_{the} \cos\theta \cong 1 - \omega_{ce} / ck $.  

For the TE mode, let us consider  semi-classical electrons first.  
With the resonance condition $\beta_r \cos\theta = 1 - \omega_{ce} / ck $,
we obtain  $\Omega_1 = ck(\cos^2\theta -1)$ and 
$\Omega_2 = ck (1-\beta_{\perp}^2 /2  - \beta_r \cos\theta )$ from Eq~(\ref{eq:te}). 
The instability criterion ($\Gamma > 0$)  is
\begin{equation} 
  \beta_r > \frac{1 - \beta_{the}^2 }{ 1-\cos^2\theta + \cos\theta} \mathrm{.}
\end{equation}
The maximum growth rate, ignoring the gyro-damping term, 
is given similar to Eq.~(\ref{eq:max}). 

For the TE mode of the fully relativistic electrons, we consider only when $\theta = \pi/2$. 
The resonance surface is given as $S = 1 - \omega_{ce} / \gamma \omega $ 
so that $\omega = \omega_{ce} / \gamma $ and 
$|\mathbf{\nabla_{\beta}} S|  = (\omega_{ce} / \omega) \gamma \beta = \gamma^2 \beta$. Note also that 
$\Omega_1 = - \omega $ and $\Omega_2 =  (1-\beta_{\perp}^2/2) ck $ from Eq.~(\ref{eq:te}). 
A similar analysis to Eq.~(\ref{eq:inst2}) can be used, and  
the instability criterion ($\Gamma > 1$)  becomes
\begin{equation} 
\int_{S=0} d^3\mathbf{\beta } \left[  \Omega_1\frac{ \mathbf{\nabla_{\beta}} S \cdot\mathbf{ \nabla_{\beta}}f}{|\nabla S(\mathbf{\beta})|^2 }  \frac{\beta_{\perp}^2}{2}\right] > \int_{S=0} d^3\mathbf{\beta } \left[ f (1-\frac{\beta_{\perp}^2}{2}) \omega \right]\mathrm{.} \label{eq:inst3}
\end{equation} 
Note that  $ \mathbf{\nabla_{\beta}} S \cdot\mathbf{ \nabla_{\beta}} f < 0 $ for a positive gyro-lasing term since $\Omega_1 <0$.  
Assuming $\beta \cong 1$, Eq.~(\ref{eq:inst3}) and is simplified to 
\begin{equation} 
\frac{ |\mathbf{\nabla_{\beta}} f|}{f} > \gamma^2 \mathrm{.}\label{eq:inst8}
\end{equation}  
In the case of the Maxwellian plasma, 
$|\mathbf{\nabla_{\beta}} f_M|/f_M = (3\gamma^2 - \lambda \gamma^3)\beta $, where $\lambda =  m_e c^2 / T_e$. 
If $\lambda \gamma < 2 $, Eq.~(\ref{eq:inst8}) and  
  $ \mathbf{\nabla_{\beta}} S \cdot\mathbf{ \nabla_{\beta}} f < 0 $ are satisfied. 
Assuming $\beta \cong 1$, it is estimated that
\begin{equation}
\Gamma \cong\frac{\pi}{2} \frac{\omega_{pe}^2}{c^2k^2} ck \int_{S=0} \frac{1}{\gamma^3} \frac{df}{d\beta}   d \mathbf{x}^3  \mathrm{.}
\end{equation}
 Note that $df / d\beta $ can be as large as $\gamma^4$,
and the maximum growth rate is given as 
\begin{equation} 
\Gamma_{max} \cong \frac{\pi}{2} \frac{1}{\zeta} \frac{\omega_{pe}^2}{c^2k^2} \omega \int_{S=0} \gamma d^3 \mathbf{\beta}  \mathrm{.} \label{eq:max4}
\end{equation}

The comparison between the case when $\theta = 0$ and when $\theta \neq 0$ is in order. 
For the case when $\theta = 0$,  $\Omega_2$ vanishes  if $\omega=ck$, 
thus the instability can occur only when $\omega \neq ck$;   
this  requires  a very high electron density 
for gamma rays or hard x-rays. 
On the other hands, when $\theta\neq 0$, 
an explosive instability is possible even when $\omega = ck \gg \omega_{pe}$.
 Therefore, 
 the case when $\theta \neq 0$ is more probable than the case when $\theta = 0$.  

\section{Gamma Ray Burst}
The  maximum growth rate of the instability in the classical plasma is given in Eq.~(\ref{eq:max}),
and the one in the fully relativistic plasmas is in Eqs.~(\ref{eq:max2}) and (\ref{eq:max4}). 
For non-Maxwellian plasmas, the growth rate 
should be  derived based on Eqs.~(\ref{eq:landau2})
and (\ref{eq:landau3}).
In this section, it is assumed that there is  sufficient free energy in the plasma so that 
the concern on the Gardner's constraint is not relevant~\citep{gardner, gardner2, sonmaxwell}.   

Regardless of whether the plasma is Maxwellian or not, 
 the maximum instability growth rate can be  summarized as 
\begin{equation} 
\Gamma_{max}\cong  \frac{\omega_{\mathrm{pe}}^2}{ (ck)^2}g(\gamma) \omega \mathrm{,} \label{eq:max5}
\end{equation}
where $ g(\gamma)\cong 1$.
   The energy loss rate via the cyclotron radiation is given as 
\begin{equation} 
 \Gamma_{ci} = \frac{2 P}{ m v_{\perp}^2} =  \frac{4}{3} \frac{ \gamma^ 2 k e^2}{m_e c^2} \omega \mathrm{,} 
\label{eq:cy}
\end{equation}
where $P$ is the loss power and  $\Gamma_{ci}$ is the ratio of the loss rate to the electron perpendicular kinetic energy. 
The condition for the instability growth rate to exceed that of the cyclotron radiation is
\begin{equation} 
\Gamma_{max} >  \Gamma_{ci}\mathrm{,} \label{eq:dom}
\end{equation}
where $\Gamma_{max}$ is given in Eq.~(\ref{eq:max}) or (\ref{eq:max5}). 
Note that the left-hand side of Eq.~(\ref{eq:dom}) is proportional to the electron density,
while the other side is not; 
the instability dominates the cyclotron radiation for higher densities.
For the non-relativistic electrons, the ratio $\Gamma_{max}/\Gamma_{ci}$ is given as
$\cong 1.79 \times (n_e/k^3) $. 
For the relativistic electrons, it is given as $ \Gamma_{max}/ \Gamma_{ci} \cong n_eg(\gamma)/k^3\gamma^2 $.

Let us define  the reference frame  \textbf{$S_0$}
where the electron average drift is zero.
Let us assume that  the resonant velocity    $v_r  = (c -\omega_{ce})/k \ll c$
in the \textbf{$S_0$}-frame. 
For a non-relativistic Maxwellian plasma in the \textbf{$S_0$}-frame, 
the density satisfying the ratio $\Gamma_{max} / \Gamma_{ci}=1$ 
for 0.01 keV (10 keV) photon is  $n_e = 10^{16} \ \mathrm{cm^{-3}}$ ($10^{25} \ \mathrm{cm^{-3}}$). 
  The photon with the energy $ \hbar ck $ in the \textbf{$S_0$}-frame is observed as a photon with the energy $\gamma_0 \hbar ck $ in the observer's frame, 
 where the $\gamma_0$ is the Lorentz factor between the \textbf{$S_0$}-frame and the observer's frame.
Noting that the relativistic factor $\gamma_0$ 
is usually between 10 and 1000 according to the general literature on the gamma ray burst~\citep{Nakar}, 
the photon of energy  $ 10 \ \mathrm{keV} <  \gamma_0 \hbar ck < 1 \ \mathrm{MeV} $ in the observer's frame corresponds to the photons of energy 
$ 0.01 \ \mathrm{keV} <  \gamma_0 \hbar ck < 100 \ \mathrm{keV} $ in the  \textbf{$S_0$}-frame, 
whose cyclotron frequency corresponds to 
 the magnetic field of $10^9$ to $10^{13}$ gauss.

\section{The instability theory versus the incoherent cyclotron radiation theory}
The theoretical consideration in our study has the following distinctive features, 
compared to the conventional cyclotron radiation theory:

($i$) 
The fastest growing E\&M mode would be the one parallel (or perpendicular) to the magnetic field. 
The intensity of a collective E\&M mode is proportional to $\exp( \gamma t \cos\theta)$ when $\theta = 0 $
and to $\exp( \gamma t \cos(\theta-\pi/2))$ when $\theta = \pi/2$,
so that the peak angle is narrow when $\gamma t \gg 1$. 
Therefore, in a certain direction, 
 more intense photons would be observed than predicted from the conventional incoherent cyclotron radiation,  whose  intensity is proportional to  $\cos^2\theta $. 
The actual energy power requirement for the gamma ray burst
might be less than that of the conventional theory~\citep{Nakar}.

($ii$) 
As the electrons emit the photons via the instability, 
the temperature becomes further anisotropic, rendering
 the E\&M wave perpendicular to the magnetic
field unstable to the Weibel instability. 
In turn, the Weibel instability mitigates the temperature anisotropy,  
emitting low-frequency photons  in the perpendicular direction.
  

($iii$)
Our analysis suggests that the gamma ray burst could 
be originated from a more compact and dense object than conventionally believed.  
In a highly dense plasma considered here, 
an incoherent photon has a short mean free path due to the gyro-damping
as well as  the Thomson scattering, while  
 the collective photons could overcome these dampings through the gyro-lasing. 
In other words, 
the compact and dense plasma could be  optically thin for the coherent photons,
while optically thick for the incoherent photons. 

To be more specific, let us consider the Thomson scattering. The decay rate is given as 
$\Gamma_T = n_e \sigma_T c \cong n_e r_e^2 c$, 
where $r_e = e^2/m_ec^2$ is the classical electron radius. 
From Eq.~(\ref{eq:max5}), the ratio $\Gamma_{\mathrm{max}} / \Gamma_T $ is given as  $\Gamma_{\mathrm{max}} / \Gamma_T \cong (1/ k r_e ) $. 
Then, the Thomson scattering is weaker than the gyro-lasing if the photon wavelength is 
longer than the classical electron radius, whose photon energy roughly corresponds to 100 MeV,
which is the case in our scenario.  
This suggests that 
the photon can get amplified, overcoming the Thomson scattering when 
the gyro-lasing from the instability  is strong. 
However, as the instability gets weaker,  the Thomson  scattering becomes eventually
the dominant decay mechanism of the photons, where the  photon mean free path is $l \cong (1/n_e r_s^2) $. 
For example, 
for $n_e = 10^{18} / \mathrm{cc}$,  $l\cong 10^6 \  \mathrm{cm} $.  
For a given object,  the collective photons could be excited  inside the object and sustained via the gyro-lasing. 
But, if there is no instability on the surface of the object, the photons will decay due to
the Thomson scattering within rather short mean free path; in order for the photons to escape from the object,
it is necessary that the gyro-lasing from the instability exists  on the surface of the object.
The delicate interplay between the Thomson scattering and the gyro-lasing in the process of the collective photons escaping from an object is beyond the scope of this paper.

($iv$)
The growth rate of the instability is sensitive to the slope of the distribution function
at the resonance.
Consider a shock region where two plasmas of different drifts violently encounter.
It is plausible that  the parallel and perpendicular electron temperatures 
are comparable but the distribution could have   two sharp humps in the parallel direction, at which case 
the instability   could be strong.

($v$)
Finally, our theory provides  an escape  scenario 
of photons from dense plasmas  in  a changing  magnetic field.  
A typical example would be the decreasing magnetic field 
(e.g., the foot point of the solar corona).
Consider the TE mode when  a photon propagates with $\theta=\pi/2$ 
from $r=0$ and  the magnetic field decreasing with the increasing $r$ (or $\mathbf{B} = B(r) \hat{z} $ with $B(r_1) < B(r_2) $ if $r_1 > r_2$).
From the resonance condition $\omega = \omega_{ce} /\gamma $, the relativistic factor $\gamma$
of a resonant electron should decrease with the decreasing $\omega_{ce} $.  
Assuming the electron temperature remains the same along the photon propagation,
the gyro-lasing term stays positive if  $\lambda \gamma < 2 $ at $r=0$ (as analyzed in Sec. 4).  
Then,  collective photons are radiated primarily 
in the direction of (perpendicular to) the magnetic field as    
 photons from the incoherent cyclotron radiation  suffer  a severe damping and 
 photons  not parallel with (or perpendicular to) the magnetic field 
experience  a weaker gyro-lasing term than the photons  parallel (or perpendicular) to the magnetic field.
    

\section{Summary}
A scenario of the gamma ray burst,
based on the recent radiation theory~\citep{sonmaxwell}, is proposed and examined.
The previous analysis~\citep{sonmaxwell} is generalized to an arbitrary angle. 
The estimation shows  that  the coherent burst of 10 keV to 1 MeV photons  
is plausible in a relativistic plasma when $\gamma_0 = 10\sim  1000$ and 
the electron density is higher than some critical value, $n_e > 10^{18} \sim 10^{26}  \ \mathrm{cm}^{-3}$. 
It is shown that a rather compact dense object 
with less available energy could cause the short gamma ray burst and  
that the observed gamma rays would be coherent rather than incoherent.
In addition, the coherent photons of lower frequency comparable to
the plasma frequency might be observed due to the Weibel instability. 
This is particularly relevant to the case when $\theta = \pi/2$
because the low frequency coherent photons (high frequency coherent photons)
from the Weibel instability (from the instability studied here) can be observed simultaneously.
The above conjecture  might be useful in verifying whether the scenario proposed here
would account for some of the short gamma ray burst events observed in the satellites~\citep{Nakar}.

While our estimation is rather focused on the gamma and hard x-rays, 
a similar mechanism would manifest in generating soft x-rays in the inertial
confinement fusion plasma~\citep{tabak}. 
The electron beam of $\gamma > 10 \sim 100$ and the magnetic field of $10^8$ gauss
can be readily generated in laboratories.
Even a magnetic field of $10^9$ gauss might be possible~\citep{sonprl}.
Then, the photon generated from the instability may have energy between 10 eV and 1 keV.
Complications would be the electron quantum diffraction effect and
the degeneracy~\citep{sonpla, sonprl, sonlandau}.
The plausibility study is in progress.

\bibliography{tera2}

\end{document}